\documentstyle[psfig,conf_iap]{article}
\def\dg{\mbox{$^\circ$}}		

\begin{document}
\heading{%
The Sloan Digital Sky Survey
} 
\par\medskip\noindent
\author{%
Jon Loveday$^1$, Jeff Pier$^2$ 
(for the SDSS Collaboration)
}
\address{%
Astronomy \& Astrophysics Department, University of Chicago, 5640 S Ellis Ave, Chicago,
IL 60637, USA.
}
\address{%
US Naval Observatory, PO Box 1149, Flagstaff, AZ 86002-1149, USA.
}
%
%
\begin{abstract}
The Sloan Digital Sky Survey (SDSS) will carry out a digital photometric
and spectroscopic survey over $\pi$ steradians in the northern Galactic cap.
An array of CCD detectors used in drift-scan mode will image the sky
in five passbands to a limiting magnitude of $r' \sim 23$.
Selected from the imaging survey, $10^6$ galaxies, $10^5$ quasars and 
selected samples of stars will be observed spectroscopically.
In addition, a smaller (225 deg$^2$), deeper,
southern survey will reach $\sim 2.0$ magnitudes fainter
and will contain a wealth of information about variable sources, supernovae
and proper motions.
We describe the current status of the survey, which recently saw first light,
and its prospects for constraining cosmological models.
\end{abstract}
\section{Survey Overview}

The SDSS consists of two surveys, one photometric and one spectroscopic.
To complete a digital survey over a large fraction of the sky within a finite 
amount of time, it is necessary to conduct wide-field imaging and multi-object
spectroscopy.  To meet this
need, a wide-field telescope, imaging camera and multi-fibre spectrographs
were designed and built specifically
for this purpose.

The alt-az telescope uses a modified Ritchey-Chr\'etien design
\cite{wmgk}, with a primary aperture of 2.5m and a focal ratio
of f/5 to produce a flat field of 3\dg with a plate scale of 16.51
arcsec/mm.  It is situated at Apache Point Observatory, near Sunspot, New
Mexico, at a height of 2,800m.  The telescope is housed in an enclosure
which rolls off for observing, and is encased in a co-rotating baffle which
protects it from wind disturbances and stray light.
This unique design allows the telescope to remain free of dome-induced seeing.
First-light images and recent photographs of the site and telescope
can be found at {\tt http://www.sdss.org/}.
Technical details about the survey can be obtained from 
{\tt http://www.astro.princeton.edu/BBOOK/}.

\subsection{Imaging Survey}

The photometric imaging survey will produce a database of roughly 10$^8$ 
galaxies and 10$^8$ stellar objects, with accurate ($\leq$ 0.10 arcsec) 
astrometry, 5-colour ($u'$, $g'$, $r'$, $i'$ and $z'$) photometry, 
and object classification parameters.  This database
will become a public archive.

The imaging camera \cite{JEG98} 
consists of 54
CCDs in eight dewars and spans $2.3\dg$ on the sky.  Thirty of these CCDs
are the main imaging/photometric devices, each a SITe (Scientific
Imaging Technologies, formerly Tektronix) device with 2048x2048  24$\mu$
pixels.  They are arranged in six column dewars of 5 CCDs each, one CCD for
each filter bandpass in each column.  The camera will be
operated in TDI (Time Delay and Integrate), or scanning, mode for which
the telescope
will be driven at a rate synchronous with the charge transfer rate of
the CCDs.  Objects on the sky will drift down columns of the CCDs so
that nearly simultaneous 5-colour photometry will be obtained with no
dead-time.
The effective integration time,
{\it i.e.} the time any part of the sky spends on each chip, is 55 seconds
at the chosen (sidereal) scanning rate, which will result in a limiting
magnitude of $r' \sim 23$.

The SDSS photometric system \cite{FUK} has been 
specifically designed for this survey and covers the near-UV to near-IR range
($\sim 3000$--$10000$\AA).
In order to provide photometric calibration while the imaging camera is
scanning, a second, dedicated {\it Monitor Telescope} will be in 
operation, observing photometric standard stars and creating photometrically
calibrated ``secondary patches'' which lie within the main telescope's
scan.  These calibration patches will then be used to transfer the
primary photometric calibration to objects detected with the 2.5m telescope
and imaging camera.

The other 24 CCDs in two additional dewars are also SITe chips of width
2048  24$\mu$ pixels,
but they have only 400 rows in the scanning direction.  These dewars are
oriented perpendicular to the photometric dewars and one lies across the
top of the columns, the other across the bottom.  Two of these CCDs
(one in each dewar)
will be used to determine changes in focus.  The other 22 CCDs will
reach brighter magnitudes before saturating and are
used to tie observations to an astrometric reference frame.

The northern survey area is centered near the North
Galactic Pole and it lies within a nearly elliptical shape $130\dg$ E-W
by $110\dg$ N-S chosen to minimize galactic foreground extinction.
All scans will be conducted along great circles in order to
minimize the transit time differences across the camera array.  
There are 45 great circles (``stripes'') in the northern survey region 
separated by 2.5\dg.
Each stripe will be scanned
twice, with an offset perpendicular to the scan direction in order
to interlace the photometric columns.  A completed stripe width will
slightly exceed $2.5\dg$ and thus there will be a small amount of overlap
to allow for telescope mis-tracking and to provide
multiple observations of some fraction of the sky for quality assurance
analysis.  
The total stripe
length will require a minimum of 650 hours of
pristine photometric and seeing conditions to scan at a sidereal rate.
Based upon historical records of observing conditions at APO, we
are allowing 5 years to complete the survey.

\subsection{Spectroscopic Survey}

The spectroscopic survey will observe spectra for $10^6$ galaxies,
10$^5$ QSOs and $10^5$ stars.
In order to obtain the spectra of over 10$^6$ objects in a survey
covering 10$^4$ square degrees, we must get spectra of about 100
objects per square degree.  Although some overlap of fields is
inevitable, we would like to keep this overlap to a minimum for reasons
of efficiency and cost.  Hence we need
to obtain several hundred spectra per 3\dg diameter spectroscopic field.

To accommodate this requirement, two identical multi-fibre spectrographs
have been built which will each be fed by 320 fibres.  The spectrographs
cover 3900--9100\AA\  with a resolution of $\lambda/\Delta\lambda \sim$ 1800,
or 167 km s$^{-1}$.  Each spectrograph has two cameras, one optimized for
the red and the other for the blue.  Each camera has as its detector
a 2048x2048 CCD with 24$\mu$ pixels.

The 180$\mu$ ($3''$ on the sky) fibres are located in the focal plane by 
plugging them by hand into aluminum plates which are precisely drilled for
each field based upon the astrometric solution obtained from the imaging
data. To avoid mechanical interference, individual fibres can be placed
no closer than $55''$ to one another.
The plates and fibres are held in the focal plane, and coupled with
the spectrographs, by one of 9 identical rigid assemblies called cartridges.
Since all of the cartridges can be pre-plugged during the day, 5,760 spectra
can be obtained during a long night without re-plugging.  A mapping procedure
will be invoked after plugging each cartridge that will tag each fibre
to the appropriate object on the sky.

To obtain redshifts for galaxies to a limiting magnitude of
$r' \sim 18$ will require exposure times of about 45 minutes,
consisting of three 15 minute exposures.
Each field should take about an hour, including calibration
(flat field and comparison lamp) exposures and allowing for telescope
pointing and the exchange of fibre cartridges.

\subsection{Data Processing}

All of the raw data from the photometric CCDs are archived.
The frames are first read to disk, then written to DLT tape.  
Over 16\ Gb per hour are generated from the
photometric chips.  When observing in spectroscopic mode, 
the amount of data generated seems trivial in
comparison (about 6 exposures per hour for each of two cameras for each
of two spectrographs, or 24 8Mb frames per hour).

All data tapes are shipped by overnight express courier to Fermi National
Accelerator Laboratory (near Chicago) where the software data reduction
pipelines will be run.
The goal is to be able to turn the imaging data around within a few
days, so that one dark run's worth of data will be processed before
the next dark run begins, allowing objects to be targeted for spectroscopy.
The data flow serially through several pipelines to identify, measure and
extract astronomical images and to
apply photometric and astrometric calibrations.
Once a significant area of sky has been imaged, a {\em target selection}
procedure will be run in order to select objects for followup spectroscopy.
Spectroscopic reduction will be automated, with the aim of obtaining
redshifts for 99\% of targeted objects without human intervention.
The pipelines are integrated into a specially-written environment
known as Dervish,
and the reduced data will be written into an object-oriented database.

\subsection{Spectroscopic Samples}

There will be several distinct spectroscopic samples observed by the survey.
In a survey of this magnitude, it is important that the selection 
criteria for each class remain fixed throughout the duration of the survey.
Therefore, we plan to spend a whole year obtaining
test data with the survey instruments and refining the spectroscopic selection
criteria in light of our test data.
Then, once the survey proper has commenced, these criteria will be
``frozen in'' for the duration of the survey.
The numbers discussed below are therefore only preliminary, and we expect
them to change slightly during the test year.

The {\bf main galaxy sample} will consist of $\sim 900,000$ galaxies selected
by Petrosian magnitude in the $r'$ band, $r' \lsim 18$.
Simulations have shown that the Petrosian magnitude,
which is based on an aperture defined by the ratio of light within an annulus 
to total light inside that radius, provides probably the least biased
and most stable estimate of total magnitude.
There will also be a surface-brightness limit, so that we do not
waste fibres on galaxies of too low surface brightness to give a reasonable
spectrum.
This galaxy sample will have a median redshift $\langle z \rangle \approx 0.1$.

We plan to observe an additional $\sim 100,000$
{\bf luminous red galaxies} to $r' \lsim 19.5$.
Given photometry in the five survey bands, redshifts can be estimated
for the reddest galaxies to $\Delta z \approx 0.02$ or better \cite{c95},
and so one can also predict their luminosity quite accurately.
Selecting luminous red galaxies, many of which will be cD galaxies in cluster 
cores, provides a valuable supplement to the main
galaxy sample since 1) they will have distinctive spectral features,
allowing a redshift to be measured up to 1.5 mag fainter than the main sample,
and 2) they will form an approximately volume-limited sample with a
median redshift $\langle z \rangle \approx 0.5$.
They will thus provide an extremely powerful sample for studying
clustering on the largest scales and galaxy evolution.

{\bf Quasar} candidates will be selected by making cuts
in multi-colour space and from the FIRST radio catalogue \cite{bwh95},
with the aim of observing $\sim 100,000$ quasars.
This sample will be orders of magnitude larger than any existing quasar
catalogue, and will be invaluable for quasar luminosity function, evolution
and clustering studies
as well as providing sources for followup absorption-line observations.

In addition to the above three classes of spectroscopic sources, which are
designed to provide {\em statistically complete} samples, we will also 
obtain spectra
for many thousands of {\bf stars} and for various {\bf serendipitous}
objects.
The latter class will include objects of unusual colour or morphology
which do not fit into the earlier classes, plus unusual objects found
by other surveys and in other wavebands.

\section{Current Status}

The 2.5m telescope and imaging camera are complete and in place.
First light with the imaging camera was obtained on 9 May 1998
during bright time and without the baffles; subsequently two imaging runs
were made during dark time and with the baffles in place.
The image size was $0.95$--$1.1''$ for the second dark run.

The spectrographs are both at the site and all optics have been
completed and coated.  One spectrograph has been fully assembled and
some test spectra taken.  A fully assembled fibre cartridge is 
ready and all the others are ready for assembly.  
The full complement of over 6,000 science fibres needed for
the survey have been accepted and tested with a mean throughput of 92.0\%.
Test plug plates have been drilled with all positions well within
tolerances.  The various pieces of equipment for storing, handling, and
transporting the cartridges are all in place.
Once the telescope control system is in operation later this year, 
we will be able to take spectra on the sky.

All of the data reduction-pipelines are written, with 
ongoing work on minor bug-fixes, speed-ups and integration of the entire
data processing system.
The imaging/photometric reduction pipelines are being exercised with
the data taken this May and June by the survey imaging camera.
Tests are being carried out on the spectroscopic reduction pipeline using
simulated data.

%

The intent of this project is to make the survey data available to the
astronomical community in a timely fashion. We currently plan to
distribute the data from the first two years of the survey no later
than two years after it is taken, and the full survey no later than
two years after it is finished.
The first partial release may or may
not be in its final form, depending on our ability to calibrate it
fully at the time of the release. The same remarks apply to the
release of the full data set, but we expect the calibration effort to
be finished before that release.

\section{Some Cosmological Highlights}

\begin{figure}
\centerline{\vbox{
\vspace{-0.8cm}
\psfig{figure=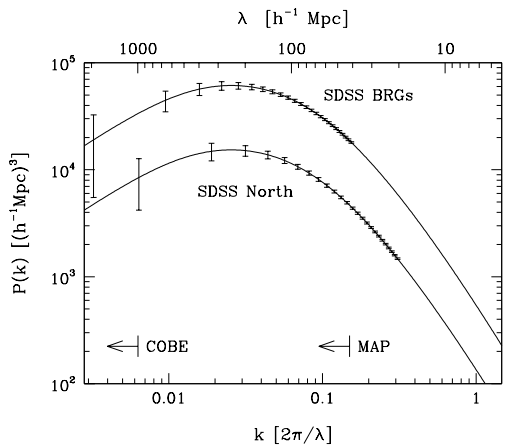,height=7.cm}
\vspace{-0.8cm}
}}
\caption[]{Model CDM power spectra and expected $1\sigma$ errors for the SDSS
Northern galaxy survey and for the luminous red galaxies (BRGs).
The latter are assumed to be biased by a factor of two with
respect to the former. Also shown are the smallest comoving
wavelength scales accessible to COBE and to the upcoming Microwave
Anisotropy Probe (MAP).}
\label{fig:P_k}
\vspace{-0.2cm}
\end{figure}

The main impetus for carrying out the Sloan survey is to provide definitive
measures of the local ($z \lsim 0.5$) large-scale structure in the Universe.
The huge volume of the SDSS redshift survey will enable reliable
estimates of the galaxy power spectrum to $\sim$ Gpc scales,
thus allowing one to constrain the shape of the primordial spectrum predicted
by linear perturbation theory (Figure~\ref{fig:P_k}).
Comparison with CMB anisotropy data, particularly the upcoming
MAP and Planck surveys,
will allow a direct measurement of galaxy bias $b$ over a wide range of scales.
Measurements of higher-order galaxy clustering (Frieman, these proceedings)
and redshift space distortions will provide further constraints on the density
parameter $\Omega_0$ and $b$.

Clues to the physics of galaxy formation will be provided by studying the
luminosity function and clustering of galaxies separated by colour,
morphology and other intrinsic galaxy properties.
Photometric redshifts \cite{c95} will make the Sloan dataset an extremely 
powerful one for studying galaxy evolution over a range of redshifts
beyond that reached by the spectroscopic survey.

We also plan to search for low surface brightness (LSB) galaxies in the Sloan
imaging data.
The drift-scan observing mode means that detection of LSB objects is limited
by photon statistics rather than flat-fielding errors, and we expect to
detect galaxies with a central surface brightness as low as 
$\mu_0 \approx 27.5$ mag arcsec$^{-2}$ in the $r'$ band.
One of the survey products will be a $4 \times 4$ binned sky map, from which
we will be able to find galaxies to a surface brightness of yet four times less.


\acknowledgements{
The SDSS Collaboration includes participants from The University of
Chicago, Fermilab, the Institute for Advanced Study, the Japan
Participation Group, The Johns Hopkins University, Princeton University,
the United States Naval Observatory and the University of Washington.
Apache Point Observatory, site of the Sloan Digital Sky Survey, is owned
by the Astrophysical Research Consortium, a nonprofit corporation
consisting of seven research institutions, and is operated by New Mexico
State University. Funding for the project has been provided by
the Alfred P. Sloan Foundation, SDSS member institutions, the
National Science Foundation and the U.S. Department of Energy.}

\end{document}